\documentclass[aps,prl,twocolumn]{revtex4-2}

\usepackage{amsmath,upgreek}
\usepackage{graphicx}
\usepackage{xcolor}
\usepackage[colorlinks,urlcolor=blue]{hyperref}
\usepackage[normalem]{ulem}
\usepackage{cancel} 
\usepackage{comment}
\usepackage{nicefrac}
\usepackage{dcolumn}

\begin{document}

\title[ ]{High-precision Penning-trap spectroscopy of the ground-state spin structure of HD$^+$}

\author{Charlotte M. König}
\author{Matthew Bohman}
\email{matthew.bohman@mpi-hd.mpg.de}
\author{Fabian Hei\ss{}e}
\author{Jonathan Morgner}
\author{Tim Sailer}
\author{Bingsheng Tu}
\altaffiliation{Present address: Institute for Modern Physics, Fudan University, Shanghai 200433, China}
\author{Klaus Blaum}
\author{Sven Sturm}
\affiliation{Max-Planck-Institut für Kernphysik, Saupfercheckweg 1, Heidelberg, 69117, Germany}

\author{Dimitar Bakalov}
\affiliation{Institute for Nuclear Research and Nuclear Energy, Bulgarian Academy of Sciences, Tsarigradsko Chaussée 72, Sofia, 1784, Bulgaria}

\author{Hugo D. Nogueira}
\author{Jean-Philippe Karr}
\altaffiliation{Also at Université d'Evry-Val d' Essonne, Université Paris-Saclay, Evry, 91000, France}
\affiliation{Laboratoire Kastler Brossel, Sorbonne Université, CNRS, ENS, Université PSL, Collège de France, Paris, 75005, France}

\author{Ossama Kullie}
\affiliation{Theoretical Physics, Institute for Physics, Department of Mathematics and Natural Science, University of Kassel, Kassel, Germany}

\author{Stephan Schiller}
\affiliation{Institut für Experimentalphysik, Heinrich-Heine-Universität Düsseldorf, Düsseldorf, 40225, Germany}

\newpage
    \begin{abstract}
    We present high-precision spectroscopy of the ground-state hyperfine structure of HD$^+$ at 4~T. We determine 
    the bound-electron $g$ factor, $g_{e,\mathrm{bound}} = -2.002\,278\,540\,96(40)$, to a relative uncertainty of $2\times$10$^{-10}$, the most precise determination of a bound-electron $g$ factor of a molecular ion to date. The experimental value agrees with recently developed ab initio theory that now includes quantum-electrodynamical effects up to order $\alpha^5$ and has reduced the theoretical uncertainty by three orders of magnitude [O. Kullie \textit{et al.}, Phys. Rev. A 112 052813 (2025)]. In addition, we extract the scalar spin-spin interaction coefficients $E_4$~=~925\,395.758(41)$\,$kHz (electron-proton) and $E_5$~=~142\,287.821(22)$\,$kHz (electron-deuteron), which show a moderate tension with another state-of-the-art theoretical prediction [M. Haidar \textit{et al.}, Phys. Rev. A 106 042815 (2022)].
    \end{abstract}

\maketitle

The search for physics beyond the Standard Model requires isolating potentially minute deviations from the overwhelmingly dominant effects of known physics. This is certainly the case for precision experiments at low energy that compare measured observables to predicted values from fundamental theory, in particular quantum electrodynamics (QED), and independently determined fundamental constants. By choosing systems that are sensitive to particular interactions, such experiments become a search platform that is competitive with, and complementary to, accelerator-based high-energy experiments \cite{Saf18}. Alternatively, under the assumption that the Standard Model is correct, precise measurements and theory can be combined to determine fundamental constants to high precision \cite{Parker2018, Morel2020, Fan23, karr2025determination}.
\smallskip

Similar to one of the simplest and most studied atomic systems, hydrogen, diatomic molecular hydrogen ions (MHI) contain only a single electron and are exceptionally well-described by quantum electrodynamics (QED) theory \cite{karr2020, korobov2021}. However, as molecular systems, MHI possess additional rotational and vibrational degrees of freedom that give rise to a rich energy spectrum comprising hundreds of rovibrational states in the electronic ground state. These energies depend on fundamental constants, such as the proton-to-electron mass ratio $\nicefrac{m_p}{m_e}$, the proton charge radius $r_p$, the Rydberg constant $R_\infty$, and the fine-structure constant $\alpha$, with well-understood scaling \cite{korobov2021}. As a result, high-precision spectroscopy of MHI in conjunction with ab initio theory can be used to determine fundamental constants, test molecular QED, and probe beyond-Standard-Model (BSM) forces \cite{alighanbari2020, alighanbari2023, germann2021, karr2025determination, Alighanbari2025}.
\smallskip

Furthermore, measurements of the rovibrational transition frequencies of H$_2^+$ and its antimatter counterpart $\overline{\mathrm{H}}_2^-$ are also very attractive for a future test of CPT invariance \cite{dehmelt1995,myers2018,schenkel2024,Vargas2025,Shore2025,Koenig2025} that would provide a unique opportunity to compare the properties of the proton and antiproton along with proton-proton (p-p) and antiproton-antiproton ($\bar{\mathrm{p}}$-$\bar{\mathrm{p}}$) interactions. Compared to atomic (anti)hydrogen, such a test would be three orders of magnitude more sensitive to the proton/electron vs. antiproton/positron mass ratio difference $\nicefrac{m_p}{m_e} - \nicefrac{m_{\bar{p}}}{m_{\bar{e}}}$. While production of $\overline{\mathrm{H}}_2^-$ remains a challenge \cite{zammit2019, taylor2024, zammit2025}, recent years have seen considerable development in precision spectroscopy of H$_{2}^{+}$ and HD$^{+}$ which now enables comparisons of experiment and theory at the level of $1\times$10$^{-11}$ relative uncertainty \cite{kortunov2021, germann2021, alighanbari2023, Alighanbari2025}.
\smallskip

In this work we aim to use the techniques of single-ion precision Penning-trap experiments to perform high-precision measurements on HD$^{+}$ and extend the reach of MHI as a platform for fundamental physics studies. These techniques, namely electron spin resonance spectroscopy (ESR) enabled by image current detection and the continuous Stern-Gerlach effect (CSGE) \cite{dehmelt1986}, have recently been applied to HD$^{+}$ and provide deterministic and non-destructive state preparation and detection \cite{Koenig2025}. This experimental platform presents an opportunity for a strong improvement in experimental resolution. For context, the only previous ESR experiment \cite{loch1988} on MHI was performed nearly 40~years ago on an ensemble of ions and determined bound-electron \textit{g} factors to a relative precision of around $1\times10^{-6}$. Today, experiments using single atomic ions routinely perform measurements of bound-electron \textit{g} factors to sub-ppb or even lower uncertainty \cite{Sturm2014, morgner2023, Dickopf2024}. \smallskip

This perspective of greatly improved experimental accuracy triggered a new ab initio calculation of the bound-electron $g$ factor that improved the only, more-than-40-year-old previous calculation \cite{hegstrom1979} by more than three orders of magnitude and allows us to now perform a high-precision comparison of theory and experiment. Such comparisons, and, more generally, accurate determinations of the hyperfine structure (HFS), are crucial as HFS affects all rovibrational transition measurements in MHI and are required to extract spin-averaged frequencies from measured transition frequencies. Currently, the extraction of fundamental constants from MHI spectroscopy is limited by the theoretical uncertainty of the spin-averaged frequencies. However, once these theoretical calculations improve, precise knowledge of the HFS will be required for improved determinations of these fundamental constants. Moreover, while theory \cite{haidar2022} and experiment \cite{jefferts1969, menasian_thesis, Alighanbari2025} on the HFS of H$_2^+$ currently agree, discrepancies have been observed in HD$^+$\cite{karr2023} that call for a resolution.
\smallskip

Here, we present the results of a new determination of the ground-state HFS of HD$^{+}$ using ESR in a precision Penning-trap experiment and measure the bound-electron \textit{g} factor with orders of magnitude improved precision. To perform these measurements, we directly drive magnetic dipole transitions between the electron spin states of a single molecular ion in the \textsc{Alphatrap} cryogenic Penning-trap apparatus \cite{alphatrap2019}. In the strong 4-T magnetic field used to trap the ion, these transition frequencies (around 112\,GHz) can be measured via millimeter-wave (MW) spectroscopy to  well below 100\,Hz precision.
\smallskip

\textit{HFS Theory}---
The effective Hamiltonian$\,$\cite{bakalov2006}
\begin{align} 
    H(v=0,N=0) &= \, hE_4(0,0)(\textbf{I}_\textnormal{p}\cdot\textbf{s}_\textnormal{e})+ hE_5(0,0)(\textbf{I}_\textnormal{d}\cdot\textbf{s}_\textnormal{e}) \nonumber\\
    &-\mu_\textnormal{B}g_\textnormal{e,bound}(0,0)(\textbf{B}\cdot\textbf{s}_\textnormal{e})\nonumber\\
    &-\mu_\textnormal{B}g_\textnormal{p}(0,0)(\textbf{B}\cdot\textbf{I}_\textnormal{p}) \nonumber\\ 
    &-\mu_\textnormal{B}g_\textnormal{d}(0,0)(\textbf{B}\cdot\textbf{I}_\textnormal{d})\ \label{eq:HwithB}
\end{align}

\noindent describes the HFS of HD$^+$ in the rovibrational ground state (vibrational quantum number $v=0$ and rotational quantum number $N=0$) in the presence of an external magnetic field $\textbf{B}$. The particle-spin orientations are indicated by the (approximate) magnetic quantum numbers $m_\mathrm{s}$, $m_{\rm I,p}$, and $m_\mathit{\rm I,d}$. The Hamiltonian characterizes the coupling of the electron spin $\textbf{s}_\mathrm{e}$ to the proton spin $\textbf{I}_\mathrm{p}$ and to the deuteron spin $\textbf{I}_\mathrm{d}$. The strengths of these couplings are given by the spin-spin interaction coefficients $E_4(0,0)$ and $E_5(0,0)$, respectively, and have been calculated ab initio with uncertainty below 1$\,$ppm $\,$\cite{karr2020}. At the same time, $H$ characterizes the Zeeman interaction of the magnetic moments due to the particle spins and $\textbf{B}$. Its strength is given by the rovibrational-state-dependent $g$ factors and the magnetic field $\textbf{B}=B\textbf{e}_\textnormal{z}$ where $\textbf{e}_\textnormal{z}$ is the unit vector along the $z$-axis. Here, $\mu_\mathrm{B}=\frac{e\hbar}{2m_\mathrm{e}}$ is the Bohr magneton, $g_\mathrm{e,bound}(0,0) \approx -2$ is the $g$ factor of the bound electron,  and $g_\textnormal{p}(0,0)$, $g_\textnormal{d}(0,0)$ are the $g$ factors of bound proton and deuteron, respectively. We emphasize that all $g$ factors, along with the $E_4$ and $E_5$ coefficients, are functions of the rovibrational level $(v,N)$. We also note that the bare nuclear $g$ factors $g_\mathrm{p}$, $g_\mathrm{d}$ may be used as their contribution to the measured transition frequencies is small (for details see the Supplemental Material (SM) \cite{SM}).\smallskip

\begin{figure}[]
\centering
\includegraphics[scale=1.0]{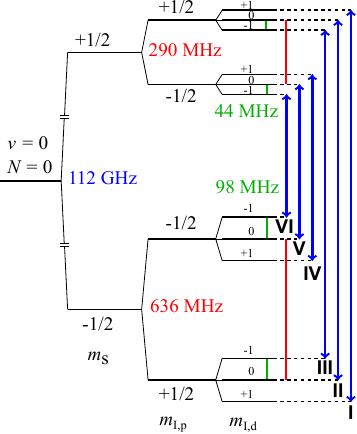}
\caption{The level scheme of HD$^+$ hyperfine structure in the rovibrational ground state at 4.02$\,$T. The blue arrows represent the six measured electron-spin-flip transitions and the red and green lines indicate the energy level splittings associated with the proton and deuteron spins, respectively. The first excited rotational state is separated from the ground state by about $1.3$\,THz.}\label{fig:level_resonances}
\end{figure}

Further corrections to $H$ are discussed in the SM \cite{SM} and are small compared to our measurement precision. The explicit Hamiltonian matrix can be found in the SM of Ref.$\,$\cite{schiller2023}. Level energies and transition frequencies are obtained by diagonalizing $H$ and are displayed in Figure$\,$\ref{fig:level_resonances}.
\smallskip

\textit{Theory of the bound-electron \textit{g} factor}--- The theoretical uncertainty of the bound-electron \textit{g} factor in diatomic MHI has been improved by more than three orders of magnitude in Ref$.$\,\cite{kullie2025precisioncalculationboundelectrong} and we summarize the results here. Using the Non-Relativistic Quantum Electrodynamics (NRQED) approach \cite{Paz_2015}, corrections to the bound-electron $g$ factor are expressed in the form of an expansion in $\alpha$, $Z\alpha$, and $m_{\mathrm{e}}/M$, where $Z$ and $M$ denote the charge and mass of a nucleus, respectively. The bound-electron $g$ factor can then be written as
\begin{equation}
g_{\mathrm{e,bound}} \left(v,\!N,\!m_N\right) = g{_\mathrm{e}} \!+\! \Delta g^{(2)} \!+\! \Delta g^{(3)} \!+\! \Delta g^{(4)} \!+\! \Delta g^{(5)} \!+\! \ldots \,, \label{eq:gfact-expansion}
\end{equation}
\noindent where $g_{\mathrm{e}}$ is the free-electron $g$ factor, and $\Delta g^{(n)}$ denotes a correction of leading order $\alpha^n$, which can be further separated into non-recoil ($\Delta g^{(n)}_{\mathrm{nonrec}}$, i.e. zero-order in $m_{\mathrm{e}}/M$) and recoil ($\Delta g^{(n)}_{\mathrm{rec}}$) contributions. With the exception of $g_{\mathrm{e}}$, all terms depend on the rovibrational and Zeeman state $\left(v,N,m_N\right)$ although, for brevity, this is not explicitly indicated.
\smallskip

Previously only the leading-order, non-recoil [$\left(Z\alpha\right)^2$] relativistic correction had been calculated \cite{hegstrom1979} while in the recent work~\cite{kullie2025precisioncalculationboundelectrong}, the theory was improved by calculating the complete $\left(Z\alpha\right)^2$ correction, including recoil~\cite{karr2021}, without using the adiabatic approximation. The new calculation also includes the non-recoil contributions of order $\alpha^3$, $\alpha^4$ and (partially) $\alpha^5$. The $\alpha\left(Z\alpha\right)^2$-order correction $\Delta g^{\left(3\right)}$ is the leading one-loop radiative correction and originates from the electron’s anomalous magnetic moment ($g_{\mathrm{e}}-2$). The next term, $\Delta g^{\left(4\right)}$,  is a pure relativistic correction of order $\left(Z\alpha\right)^4$. The relativistic $g$ factor $g_{\mathrm{rel}}$ was calculated with high precision using wavefunctions obtained by the minmax finite element method presented in Ref.$\,$\cite{kullie2022} while the relativistic corrections of orders $\left(Z\alpha\right)^4$ and above were obtained by subtracting the leading orders from the result:
\begin{equation}
\Delta g_{\mathrm{rel}}^{\left(4+\right)} = g_{\mathrm{rel}} - g_{\mathrm{e}} - \Delta g_{\mathrm{nonrec-ad}}^{(2)} \, , \label{eq:grel}
\end{equation}
where, in the right-hand side, the Dirac value $g_{\rm e} = 2$ is used and calculations are performed in the adiabatic approximation. Finally, at the $\alpha^5$ order, the largest contribution by far is the one-loop self-energy $\Delta g_{\mathrm{SE}}^{(5)}$, with a much smaller contribution from one-loop vacuum polarization. The self-energy correction is a sum of logarithmic $\alpha\left(Z\alpha\right)^4 \ln(\alpha)$ and non-logarithmic terms.\smallskip

Overall, the fractional theoretical precision of the bound-electron $g$ factor, $g_\mathrm{e,bound}(0,0)$, is $5.0\times 10^{-11}$. The largest uncertainty comes from the estimate of the nonlogarithmic part of the one-loop self-energy correction $\Delta g_{\mathrm{SE}}^{(5)}$, and the next largest are due to the uncalculated recoil correction $\Delta g_{\mathrm{rec}}^{(3)}$ at order $\alpha\left(Z\alpha\right)^2$ along with the use of the adiabatic approximation in the calculation of $\Delta g_{\mathrm{nonrec}}^{(3)}$.
\smallskip

\textit{Experimental methods}--- The trap setup is placed in a  static and homogenous magnetic field with $B = 4.02$$\,$T and is cooled to 4$\,$K by a liquid-He bath cryostat. A room-temperature beamline connects the trap to ion sources which include a Heidelberg compact electron beam ion trap (HC-EBIT)$\,$\cite{micke2018} that we use to produce HD$^{+}$ ions. A cryogenic vacuum valve separates the beamline from the trap and allows us to achieve a vacuum pressure below $10^{-16}$~mbar. Consequently, trapping lifetimes of most ions are very long, allowing clouds of ions to be stored for months without loss \cite{morgner2023, Morgner_B_like, Koenig2025}. In total, eight individual HD$^+$ ions were isolated and used for the measurements.
\smallskip

To perform precision spectroscopy of the hyperfine structure, we utilize the ``double-trap technique''$\,$\cite{haffner2000}.
As a result, our trap stack features a ``precision trap'' (PT), with an extremely homogeneous magnetic field, $\mathbf{B}$, and a highly-harmonic electric field that supports single-ion spectroscopy to high precision. Below the PT is an ``analysis trap'' (AT) that contains a strong magnetic field gradient (``magnetic bottle''), allowing us to employ the continuous Stern-Gerlach effect to determine the spin orientation $m_{\rm s}$. Furthermore, a ``reservoir trap'' allows us to  store additional ions and extract new single ions for spectroscopy if needed.
\smallskip

In a Penning trap, due to the superposition of static electric and magnetic fields, an ion has three distinct motional modes: one axial mode due to the electric potential, and two radial modes, the modified cyclotron and magnetron modes. In the case of HD$^{+}$ at \textsc{Alphatrap} these modes have frequencies $\nu_z \approx 650\,$kHz, $\nu_+\approx 20\,$MHz, and $\nu_- \approx 10\,$kHz, respectively, which are related to the free-space cyclotron frequency,
\begin{equation}
    \nu_c = \frac{1}{2\pi}\frac{q_\mathrm{ion}}{m_\mathrm{ion}}B ,
    \label{eq:vc}
\end{equation}
via the invariance theorem$\,$\cite{brown1986}:
\begin{equation}
    \nu_c=\sqrt{\nu_+^2 + \nu_z^2 + \nu_-^2} \ ,
\end{equation}
where $\nicefrac{q_\mathrm{ion}}{m_\mathrm{ion}}$ is the charge-to-mass ratio of the ion. A measurement cycle starts by initially determining the HFS quantum state ($m_\mathrm{s}$, $m_\mathrm{I,p}$, $m_\mathrm{I,d}$) of the ion in the AT by attempting to induce an electron spin flip. Note that the frequency of this transition depends on the rovibrational and hyperfine state of the ion so that a successfully induced flip of $\Delta m_\mathrm{s} = \pm 1$ (heralded by a change of the axial frequency in the magnetic bottle) uniquely identifies the complete internal quantum state \cite{Koenig2025}. The ion is then adiabatically transported to the PT where we attempt to drive an electron spin transition using much lower MW amplitude to avoid power broadening. Simultaneously, the magnetic field is determined to high precision by measuring the three motional frequencies of the ion. \smallskip

After transporting the ion back to the AT the HFS state ($m_\mathrm{s}$, $m_\mathrm{I,p}$, $m_\mathrm{I,d}$) is determined again to reveal whether the MW excitation in the PT has successfully induced an electron-spin transition. This way, the resonance is recorded in the homogeneous field of the PT with correspondingly small linewidth. A single, fully-automated measurement cycle takes approximately 35 minutes and each of the six measured transitions consists of 100--500 cycles - see End Matter (EM) for details.
\smallskip

\begin{figure*}
    \centering
    \includegraphics[scale=1.0]{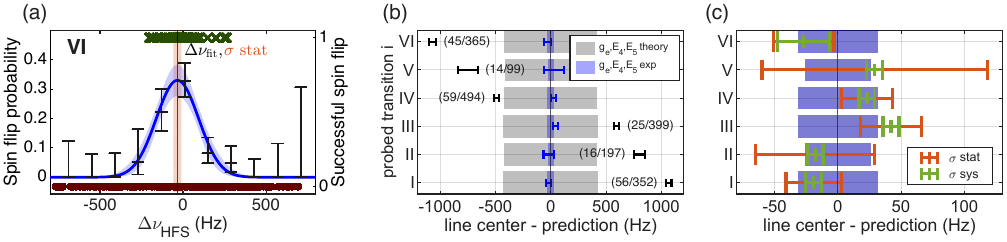}
    \caption{\textbf{(a)} Representative resonance data and fit, in this case for transition VI (shown in Fig$.$\,\ref{fig:level_resonances}). Individual spin flip attempts at a given detuning to the best-fit frequency value, $\Delta\nu_{\textnormal{HFS}}$, are marked with a dark green \textsf{X} (successful) or a dark red dot (unsuccessful). The blue curve is a Gaussian maximum-likelihood fit to the data and the blue and orange shaded areas are the confidence bands of the fitted lineshape parameters and center value, respectively. . The black binned points are provided for illustration but are not used to fit and extract values. \textbf{(b)} The fitted line center of all six resonances relative to the values calculated from the effective Hamiltonian in Eq$.$\,1 using the values of $E_4$, $E_5$, and $g_{e,\rm bound}$ from both ab initio theory (gray) and a global fit to the experimental data (blue). The shaded regions indicate the 1$\sigma$ confidence bands of these predictions and text labels indicate the number of successful and attempted spin flips. The differences in experimentally determined values of $E_4$ and $E_5$ from the ab initio calculations (see Tab.$\,$\ref{tab:results}) shift the black points right and left depending on the respective proton and deuteron orientations $m_\mathrm{I,p}$, $m_\mathrm{I,d}$ of each transition. The absence of a uni-directional shift of all transitions reflects the good agreement between experimental and theoretical values of $g_{e,\rm bound}$.  \textbf{(c)} The same as (b) showing only the results of the global fit that determines both the best fit values for $E_4$, $E_5$, and $g_{e,\rm bound}$ and the line centers. The agreement between the predicted values and the line centers shows that the data is well-described by the effective Hamiltionian with the experimentally determined hyperfine coefficients.
   }
    \label{fig:results}
\end{figure*}

\textit{Experimental Results}---
Although the room temperature ion source produces HD$^{+}$ ions in many rovibrational states, spontaneous emission causes the ions to quickly decay into the $(v=0,N=0)$ rovibrational ground state where the population largely remains due to the low blackbody radiation environment provided by the 4\,K trap chamber \cite{Koenig2025}. To measure the ground state HFS we investigated ``pure" electron-spin-flip transitions ($\Delta m_\mathrm{s} = \pm 1$), that do not change the nuclear spin orientations ($\Delta m_{\rm I,p} = \Delta m_{\rm I,d} = 0$),  marked as blue arrows in Figure$\,$\ref{fig:level_resonances}\,(a). We were able to measure MW resonances with a full width at half maximum (FWHM) of $\sim\,$300$\,$Hz. One such resonance is shown in Figure$\,$\ref{fig:results}\,(a) along with the Gaussian maximum-likelihood fit that allows us to determine the center of each line to about $8\%$ of the FWHM, corresponding to $\sim\,$2$\times$10$^{-10}$ fractional statistical uncertainty ($\sim$20$\,$Hz). Details are given in the EM.
\smallskip

In order to extract $g_\mathrm{e,bound}(0,0)$, $E_4(0,0)$, and $E_5(0,0)$ from the recorded resonances, the values of the MW drive frequencies have to be combined with the precise magnetic field $B$ at the time of the excitation, which we determine from the measured cyclotron frequency (Eq.~\ref{eq:vc}). The dominant systematic shift arises from relativistic effects due to the excited motion of the ion during the phase-sensitive measurement of the modified cyclotron frequency, $\nu_+$ \cite{Sturm2011}.
This shift is corrected for and impacts transitions I-V at the $0.25(5)\,$ppb level and transition VI at $0.90 (18)\,$ppb$\,$\cite{ketter2014}, due to a different choice of excitation strength. 
The data from all six transitions is used to determine $g_\mathrm{e,bound}(0,0)$, $E_4(0,0)$, and $E_5(0,0)$ by a combined Gaussian maximum-likelihood fit, see EM. 
The results are shown in Figure$\,$\ref{fig:results} and Tab.$\,$\ref{tab:results}. 
Other systematic shifts and corrections are at least an order of magnitude smaller and are discussed in EM and SM \cite{SM}.\smallskip

\begin{table}[]
    \centering
   
    \begin{tabular}{|l|l|l|}
    \hline 
     \textbf{a)} & \multicolumn{2}{|l|}{\small $g_\mathrm{e,bound}(0,0)$}   \\
    \hline
    \small  Hegstrom \cite{hegstrom1979}  & \multicolumn{2}{|l|}{\small  $-2.002\,278\,46(20)\,$}  \\
\hline
\small {$g_{\rm e}$ \cite{CODATA2022}} & \multicolumn{2}{|l|}{\small $-2.002\,319\,304\,360\,92(36)\,$ }\\
    \small $\Delta g_{\rm nonrec-ad}^{(2)}$ & \multicolumn{2}{|l|}{\small \hspace{1.5mm} $0.000\,040\,819\,39$} \\
    \small $\Delta g_{\rm rec-na}^{(2)}$ & \multicolumn{2}{|l|}{\small $-0.000\,000\,006\,99$} \\
    \small $\Delta g^{(3)}$ & \multicolumn{2}{|l|}{\small $-0.000\,000\,049\,05(5)$} \\
    \small $\Delta g^{(4+)}_{\rm rel}$ & \multicolumn{2}{|l|}{\small \hspace{1.5mm} $0.000\,000\,000\,52$} \\
    \small $\Delta g^{(5)}$ & \multicolumn{2}{|l|}{\small $-0.000\,000\,000\,20(9)$} \\
    \hline
    \small total theory \cite{kullie2025precisioncalculationboundelectrong} & \multicolumn{2}{|l|}{\small   $-2.002\,278\,540\,70(10)$ }  \\
 \small 
   experiment & \multicolumn{2}{|l|}{\small  $-2.002\,278\,540\,96(18)_\mathrm{st}(35)_\mathrm{sys}$ } \\
    \small  $\Delta\ (\sigma)$ & \multicolumn{2}{|l|}{\small $0.6$}   \\
    \hline
    \hline
    \textbf{b)} &  \small $E_4(0,0)$ (kHz) & \small $E_5(0,0)$ (kHz) \\
    \hline
     \small  
     \small  theory \cite{karr2023,karr2020}    & \small $925\,394.16(86)\,$ & \small $142\,287.556(84)\,$  \\
     
     experiment & \small  $925\,395.758(39)_\mathrm{st}(11)_\mathrm{sys}$   & \small  $142\,287.821(20)_\mathrm{st}(8)_\mathrm{sys}$   \\

     \small  $\Delta\ (\sigma) $  & \small $1.9$ & \small $3.1$  \\

    \hline
    \end{tabular}
     \caption{(a) \textit{ab initio} theoretical prediction and contributions to the bound-electron $g$ factor along with the measured value with the statistical (st) and systematic (sys) uncertainties. (b) \textit{ab initio} theoretical prediction and experimental results for the spin-spin interaction coefficients $E_4$ and $E_5$. In both parts (a), (b) the deviation between experiment and theory, $\Delta$, is given in units of the combined uncertainty $\sigma$.}
    \label{tab:results}
\end{table}

\textit{Discussion}---
This work presents the most precise bound-electron \textit{g}-factor measurement of any molecule to date, $g_\mathrm{e,\mathrm{bound}}(0,0) = -2.002\,278\,540\,96(40)$, corresponding to a relative uncertainty of 200$\,$ppt. The experimental value and the theoretical value agree at this level (\ref{tab:results}), where the total uncertainty is dominated by the experiment. This agreement represents an exceptional improvement compared to the previous comparison between experiment \cite{loch1988} and theory \cite{hegstrom1979}, which was unable to probe the bound-electron QED contributions (starting with $\Delta g^{(3)}$).
\smallskip

In addition, measuring several resonances in the rovibrational ground state enabled us to determine the spin-spin coupling coefficients $E_4(0,0)$ and $E_5(0,0)$ to 41$\,$Hz and 22$\,$Hz uncertainty, respectively, corresponding to a relative uncertainty of 44$\,$ppb and 151$\,$ppb. This improves on the only previous experimental determination \cite{bressel2012} by approximately 5 and 4 orders of magnitude and also surpasses the current theory \cite{karr2020} precision by factors of 20 and 4, respectively. Furthermore, our experimental uncertainties are, for the first time, small enough to be sensitive to the higher-order non-recoil QED contributions to $E_4$, a vibrational contribution that enters at the $3\times10^{-6}$ level \cite{karr2020}. We observed moderate (2$\,\sigma$ and 3$\,\sigma$) tensions to the theoretical values. The measured values of $E_4(0,0)$ and $E_5(0,0)$ differ from the theoretical values by a similar relative amount; $E_4(0,0)$ is $1.7\,$ppm larger and $E_5(0,0)$ is $1.9\,$ppm larger.
\smallskip

Recent laser spectroscopy of H$_2^+$$\,$\cite{Alighanbari2025} found experiment-theory agreement for the spin-rotation coefficient at the $\simeq0.15$\,kHz level while rovibrational spectroscopy of HD$^+$$\,$\cite{alighanbari2020,patra2020,kortunov2021,alighanbari2023} has also tested HFS calculations$\,$\cite{karr2020}. However, while some measurements in HD$^{+}$ show agreement within $\simeq1$\,kHz experimental uncertainties, others deviate by as much as 8.5 kHz \cite{patra2020, haidar2022}. This particular case corresponds to a disagreement between experiment and theory by 9$\,$$\sigma$$\,$\cite{haidar2022} and points to an emerging puzzle in HD$^{+}$.
\smallskip

Of the transitions, $(v_i,N_i) \rightarrow (v_f, N_f)$, that have been measured in HD$^{+}$, namely $(0,0) \rightarrow (0,1)$ \cite{alighanbari2020}, $(0,0) \rightarrow (1,1)$ \cite{kortunov2021}, $(0,0) \rightarrow (5,1)$ \cite{alighanbari2023}, and $(0,3) \rightarrow (9,3)$ \cite{patra2020, germann2021}, only the $(0,0) \rightarrow (5,1)$ and the $(0,3) \rightarrow (9,3)$ measurements are sensitive to $E_4$ and $E_5$ via the difference $E_{4,5}(v_f, N_f) - E_{4,5}(v_i, N_i)$. For the $(0,0)\rightarrow(5,1)$ transition, the difference of two spin component frequencies was determined \cite{alighanbari2023} and found to be in agreement with theory. If the same comparison is made using our experimentally determined values of $E_4(0,0)$ and $E_5(0,0)$ rather than the theoretically predicted values, theory and experiment remain in agreement at the 1-2 kHz level for the $(5,1)$ rovibrational state, with the uncertainty dominated by the experimental resolution \cite{alighanbari2023}. Meanwhile, the discrepancy  of 8.5 kHz on the hyperfine interval measured in the $(0,3) \rightarrow (9,3)$ transition remains unexplained. More measurements on high vibrational and rotational states will be critical for resolving this issue. See SM for further details \cite{SM}.
\smallskip

\textit{Outlook}--- The availability of high-precision theoretical predictions combined with the high experimental resolution demonstrated here establishes a general technique for unambiguously identifying the complete internal state of any MHI isotopologue. We expect this technique to be particularly useful in studies of homonuclear MHI, for which the lifetimes of excited vibrational and rotational levels are days or even longer. More specifically,  this would enable rovibrational laser spectroscopy of H$_2^{+}$ and, in the future $\bar{\mathrm{H}}_2^-$ \cite{myers2018}, from a wide variety of initially populated levels.
\smallskip

The current levels of precision of experiment and theory also suggest promising avenues for further research. As discussed, there is  a pressing need to experimentally determine the $E_4$, $E_5$ coefficients of not just the ground state, but of excited ($v > 0$, $N >0$) states, which are needed to determine spin-averaged frequencies and ultimately certain fundamental constants \cite{schiller2024,karr2025determination}. In HD$^+$, such excited rovibrational levels could be populated by laser excitation and the lifetime of these levels (of order 10\,ms) is sufficient for ESR with lifetime-limited precision. 
\smallskip

The approach demonstrated here, using single ions in a Penning trap, is highly complementary to other ongoing efforts toward improved precision measurements in MHI such as with quantum logic spectroscopy \cite{wellers2021, Holzapfel2025} and a comparison of HFS measurements at both low and high magnetic field would be of great interest.
\smallskip

More generally, the determination of the hyperfine interaction coefficients and bound-electron \textit{g} factors can directly test high-precision calculations of fundamental theory. These tests, that we show can now be performed at world-record sub-ppb precision, are exceedingly important in light of existing discrepancies between high-precision comparisons of experiment and theory and, potentially, can play a role in their resolution.
\smallskip

\textit{Acknowledgments}---
We are indebted to U.\,Rosowski (Univ.~Düsseldorf) for installing and maintaining the Hydrogen maser and GNSS receiver.
The work of St.S. and D.B. has received funding from the European Research Council (ERC) under the European Union’s Horizon 2020 research and innovation programme (grant agreement No.\,786306, ``PREMOL”).
D.B. acknowledges the partial support from Bulgarian Science Foundation Grant KP-06-N58/5.
The work of H.N. and J.-Ph.K. was part of 23FUN04 COMOMET that has received funding from the European Partnership on Metrology, co-financed by the European Union’s Horizon Europe Research and Innovation Programme and from by the Participating States. Funder ID: 10.13039/100019599. All authors from the MPI für Kernphysik, Heidelberg acknowledge funding from the Max Planck Society and the MPG-PTB-Riken Center for Time, Constants and Fundamental Symmetries.

\newpage

\section{End Matter}
\subsection{Measurement of MW Resonances}
\vspace{-10 pt}

\begin{figure*}[ht]
\centering
\includegraphics[scale=0.6]{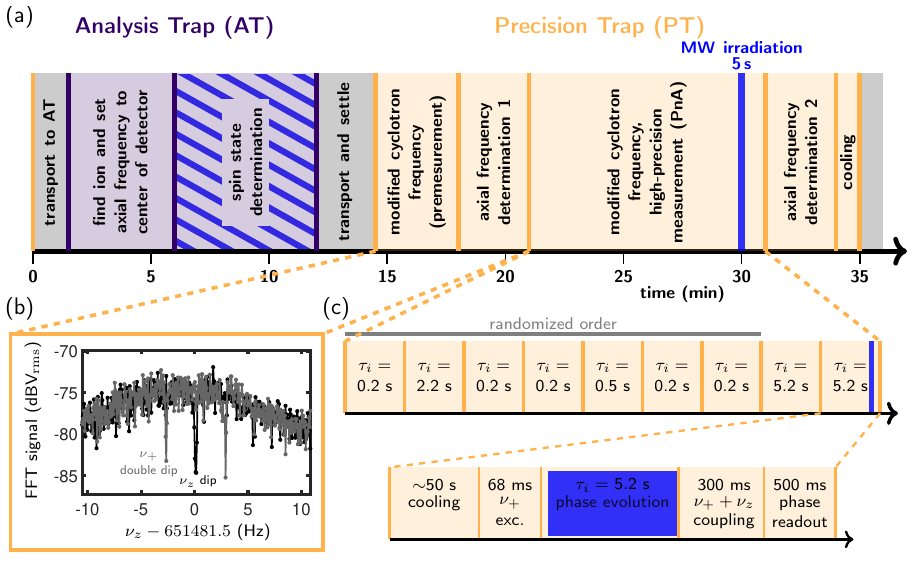}
\caption{a) Schematic of a typical measurement cycle. In the AT the internal state of the ion is determined (purple section) by observing an electron spin flip with 50$\%$ spin-flip probability - depicted via slanted blue lines. The precision measurement is performed in the PT (orange section). b) An axial dip and modified cyclotron double dip, seen on the axial resonator, both of which are used to determine the free cyclotron frequency $\nu_c$ -- see text for details. c) The measurement of the modified cyclotron frequency is depicted with a single phase-sensitive measurement sequence with a $5.2$\,s phase evolution time during which the MW is injected. Multiple phase evolution times $\tau_i$ are required for phase unwrapping and the ion is cooled for 50\,s preceding each phase measurement. Adapted from Ref$.$ \,\cite{koenig_thesis}.}\label{fig:cycle}
\end{figure*}

\textit{Measurement Sequence}---
The motion of the ion induces fA-level image currents in the trap electrodes. For the axial oscillation, this is measured as a voltage drop across a cryogenic superconducting resonant RF circuit \cite{alphatrap2019}. After thermalization, this results in a dip signal in the Fourier spectrum of the circuit output (shown in Figure\, \ref{fig:cycle}b) which can be fitted with a well-established line shape model \cite{alphatrap2019}. Both trapping regions - AT and PT - are equipped with a detection circuit. The radial modes can be measured by coupling to the axial mode via RF excitation at the sum or difference frequency resulting in a double dip in the FFT spectrum (Figure \ref{fig:cycle}b). As $\nu_+$ is the largest frequency its uncertainty dominates the final uncertainty in the $B$-field determination. Therefore, this frequency is measured more precisely with the phase sensitive \textit{Pulse-and-Amplify} (PnA) technique \cite{alphatrap2019}. In this technique the motional phase of the ion is read out after various evolution times for phase unwrapping, see Fig$.$\,\ref{fig:cycle}c.
\smallskip

In the PT the MW is irradiated for $5\,$--$\,8\,$s, resulting in an incoherent drive. The overall transmission efficiency is frequency-dependent so we adjust the MW power for each transition to avoid saturation. A typical measurement cycle is displayed in Figure \ref{fig:cycle}.
\smallskip

\textit{Data Analysis}---
Each collected data point $k$ of a transition $i= \rm I - VI$ consists of the current magnetic field value $B(k)$ obtained by a $\nu_c(k)$ measurement, the value of the simultaneously applied MW frequency $\nu_\mathrm{MW}(k)$, and the binary outcome $D(k)=0,1$ of the attempted transition. 
Diagonalizing the Hamiltonian $H$, Equation (\ref{eq:HwithB}), with $B(k)$ and arbitrary, but fixed reference values $g_\mathrm{e,bound}^{(0)}$, $E_4^{(0)}$, $E_5^{(0)}$, the predicted transition frequencies $\{\nu_{\mathrm{theo}}(k)\}_i$ are calculated.\smallskip

Note that the values $\{g_\mathrm{e,bound}^{(0)}, E_4^{(0)}, E_5^{(0)}\}$ are not necessarily the theoretical values, but can be shifted arbitrarily as long as the shifts are constant.
We chose the reference values after finding all transitions by adding offsets to the theory prediction.
Using $\nu^{\rm tot}_{\mathrm{theo}}(k)$, each data point is assigned a detuning $\Delta\nu_\mathrm{HFS}(k) = \nu_\mathrm{MW}(k) - \nu^{\rm tot}_{\mathrm{theo}}(k) $. Higher-order magnetic field effects ($\Delta\nu_{\rm ho}$) are taken into account using fixed offsets $\{\nu^{\rm tot}_{\mathrm{theo}}(k)\}_i = \{\nu_{\mathrm{theo}}(k)+\Delta\nu_{\rm ho}\}_i$ for each transition and are computed from the values in Table IV of the SM \cite{SM}.
\smallskip

The resonance data $\{\Delta\nu_\mathrm{HFS}(k),D(k)\}_k$ are first fit by a maximum-likelihood distribution with a Gaussian line shape for each transition $i= \mathrm{I-VI}$ individually, using the following function:
\begin{equation*}
    P_i(\Delta\nu_{i,\mathrm{HFS}})=A_i\,e^{-\frac{(\Delta\nu_{i,\mathrm{HFS}}-\Delta\nu_{i,\mathrm{center}})^2}{2\sigma_i^2}}\;,
\end{equation*}
where $A_i$ is the amplitude, $\Delta\nu_{i,\mathrm{center}}$ is the center of the resonance, and $\sigma_i$ is the standard deviation of the normally distributed data.
The line widths are approximately 300$\,$Hz (2.5$\,$ppb). The dominant contributions arise from the phase-imprint jitter due to the thermal motional amplitude of the ion and temporal magnetic field fluctuations during the magnetic field determination. 
The center of the lines are determined with statistical uncertainties approximately equal to $10\,\%$ of the line widths. This yields a set of six nominal line center detunings and their uncertainties $\{\Delta\nu_{\mathrm{fit}}^{i,(0)},u(\Delta\nu_{\mathrm{fit}}^{i,{(0)}})\}_{i=\textnormal{I-VI}}$.
\smallskip
Then, all data points of the six transitions are used for the determination of $g_\mathrm{e,bound}$, $E_4$, $E_5$.
The likelihood of a combined Gaussian fit of all six transitions is maximized by variation of $g_\mathrm{e,bound}$, $E_4$, and $E_5$, using the following function: 
\begin{equation}
    P_i(\Delta\nu_\mathrm{HFS})=A_i\,e^{-\frac{(\Delta\nu_\mathrm{HFS})^2}{2\sigma_i^2}}\;,
\end{equation}
where $A_i$ and $\sigma_i$ are the amplitude and standard deviation of the normally distributed data for each transition, respectively.
The fit requires the reevaluation of $\nu_{\mathrm{theo}}^{\rm tot}(k)$ and therefore, $\Delta\nu_\mathrm{HFS}(k) = \nu_\mathrm{MW}(k) - \nu_{\mathrm{theo}}^{\rm tot}(k) $ for each data point in every variation of $g_\mathrm{e,bound}$, $E_4$, $E_5$. 
The results of the combined fit are reported in Figure \ref{fig:results}.
The final $g_\mathrm{e,bound}$, $E_4$, $E_5$ are then used to obtain $\Delta\nu_{\mathrm{fit}}^{i}$ by a Gaussian maximum-likelihood fit for each transition individually.
\smallskip

As a cross check of the fit, we linearized $\nu_{\mathrm{fit}}^{i}$ around the reference values $\nu_{\mathrm{fit}}^{i,(0)}$ for which we introduce the coefficients $c_{g}^i$, $c_4^i$, $c_{5}^i$ with $c_{g}^i=(\partial \nu_{\mathrm{fit}}^{i,(0)}/\partial g_{\rm e, bound})$, etc., computed for the reference parameter values.
We compute the total squared deviation
\begin{equation}
    \sum_i (\Delta\nu_\mathrm{fit}^{i,(0)} + c_{g}^i \, \Delta g_{\rm e,bound} + c_{4}^i \, \Delta E_4 + c_{5}^i \, \Delta E_5)^2 / u(\Delta\nu_{\mathrm{fit}}^{i,(0)})^2
\end{equation}
and minimize it by varying $\Delta g_\mathrm{e,bound}$, $\Delta E_4$, $\Delta E_5$, which are defined as $g_\mathrm{e,bound}^{(0)}-g_\mathrm{e,bound}^{\rm fit}$, $E_4^{(0)}-E_4^{\rm fit}$, $E_5^{(0)}-E_5^{\rm fit}$, respectively. After minimization, $g_\mathrm{e,bound}^{\rm fit}$, $E_4^{\rm fit}$, $E_5^{\rm fit}$ represent the final result.
This procedure is admissible since the reference parameters $g_\mathrm{e,bound}$, $E_4^{(0)}$, $E_5^{(0)}$ were chosen very close to the experimentally determined ones. For both methods the same reference parameters were used. The results of the linearized approach are in agreement with the combined maximum likelihood fit described above.\smallskip

Note that the inclusion of higher order magnetic field effects via $\Delta\nu_{\rm ho}$ has minimal impact on the determined values of $E_4(0,0)$ and $E_5(0,0)$ and only slightly impacts $g_\mathrm{e,bound}$. Excluding these effects from the evaluation would change the last two digits of $g_\mathrm{e,bound}^{(0)}$ from 96 to 82.\smallskip
\vspace{-20pt}

\subsection*{Systematic Effects}
\textit{Magnetic field determination}---
The largest systematic effect occurs in the determination of $\nu_c$ and is due to special relativity, resulting in a shift of up to 0.9 ppb towards lower frequencies. For the phase sensitive measurement of $\nu_+$ the ion has to be initialized to a radial orbit larger than the thermal motion. For transitions I-V the excitation pulse was set to 68$\,$ms, corresponding to a radius ($\rho_+$) of 48$\,\mu$m and for transition VI to 135$\,$ms, corresponding to $\rho_+=96 \,\mu$m.
The resulting frequency shift for the free cyclotron frequency is $\,$\cite{ketter2014}
\begin{equation}
{\Delta\nu_c}/{\nu_c} \approx - {(2\pi\nu_+\rho_+)^2}/{2c^2} \, ,
\end{equation}
leading to 0.25 and 0.9$\,$ppb shifts, respectively. 
The uncertainty of this shift is given by the uncertainty of the PnA radius calibration ($\sim\,$10$\,\%$) resulting in a 20$\,\%$ shift uncertainty. The mass of HD$^+$ is known to 1.2$\,\times$10$^{-11}\,$\cite{rau2020} and does not limit the results here.
\smallskip
\begin{table}[]
\begin{center}
\begin{tabular}{l@{\hspace{2em}}D{.}{.}{2.4}}
\hline
 \small shift  & \multicolumn{1}{c}{\small $\Delta\nu_c/\nu_c\, (10^{-10})$}  \\
\hline
\small special relativity (I-V)\cite{ketter2014}  & -2.5(5)   \\ 
\small special relativity (VI) \cite{ketter2014}  & -9.0(18)   \\
\small  polarization \cite{rau2020,schiller2014pol}   & -0.210(1)  \\ 
\small  image charge shift \cite{schuh2019}   & -0.038(2)  \\ 
\small  $C_4$ shift    & 0.00(2)  \\ 
\small  $C_{n>4}$ shift   & 0.0000(3)  \\ 
\small  $B_2$ shift   & 0.00(2)  \\ 
\small  line shape $\nu_z$  & 0.00(1)  \\ 
\small  $\nu_-$  & 0.00(15)  \\ 
\small ion mass$\,$\cite{rau2020} & 0.00(20)\\
\hline
\small axial temperature $T_z$   & 9.2(4)\text{K}  \\
\hline
\end{tabular}
\label{tab:sys}
\caption{Systematic effects for the determination of the magnetic field.}
\end{center}
\end{table}

Further shifts and uncertainties listed in Table II are due to electric field anharmonicities ($C_4$, $C_{n>4}$) and residual magnetic field inhomogeneity ($B_2$) along with an uncertainty assigned to the dip lineshape used to fit the axial frequency $\nu_z$ and an uncertainty of the determined magnetron frequency $\nu_{-}$. All of these are smaller than the statistical uncertainty of 0.3$\,$ppb.
\smallskip

\end{document}